\documentclass[12pt]{article}
\usepackage{epsfig,amsfonts,amssymb}
\usepackage{amsmath}
\usepackage{amssymb}
\usepackage{txfonts} 
\usepackage{mathrsfs}
\usepackage{cite}
\input epsf.sty
\def\ZZZ{{\hbox{ Z\kern-1.6mm Z}}}
\def\zzz{{\hbox{z\kern-1mm z}}}

\newcommand{\eps}{\epsilon}

\newcommand{\AAA}{{\cal A}}

\newcommand{\OO}{{\cal O}}

\newcommand{\DD}{{\cal D}}

\newcommand{\LL}{{\cal L}}

\newcommand{\wt}{\widetilde}
\newcommand{\wh}{\widehat}

\newcommand{\NN}{{\cal N}}

\newcommand{\be}{\begin{equation}}
\newcommand{\ee}{\end{equation}}
\newcommand{\ben}{\begin{eqnarray}\displaystyle}
\newcommand{\een}{\end{eqnarray}}

\newcommand{\bea}[1]{\begin{eqnarray}\label{#1} }
\newcommand{\eea}{\end{eqnarray}}

\newcommand{\refb}[1]{(\ref{#1})}
\newcommand{\p}{\partial}
\newcommand{\sectiono}[1]{\section{#1}\setcounter{equation}{0}}

\def\one{{\hbox{ 1\kern-.8mm l}}}
\def\zero{{\hbox{ 0\kern-1.5mm 0}}}

\begin{document}

{}~
{}~

\vskip .6cm

{\baselineskip20pt
\begin{center}
{\Large \bf Consistent Truncation to 
Three Dimensional (Super-)gravity
} 

\end{center} }

\vskip .6cm
\medskip

\vspace*{4.0ex}

\centerline{\large \rm
Rajesh Kumar Gupta and Ashoke Sen}

\vspace*{4.0ex}

\centerline{\large \it Harish-Chandra Research Institute}

\centerline{\large \it  Chhatnag Road, Jhusi,
Allahabad 211019, INDIA}

\vspace*{1.0ex}

\centerline{E-mail:
rajesh@mri.ernet.in, sen@mri.ernet.in, ashokesen1999@gmail.com}

\vspace*{5.0ex}

\centerline{\bf Abstract} \bigskip

For a general three dimensional theory of 
(super-)gravity coupled
to arbitrary matter fields with arbitrary set of higher derivative
terms in the effective action, we give an algorithm 
for consistently truncating the
theory to a theory of pure (super-)gravity with the gravitational
sector containing only Einstein-Hilbert, 
cosmological constant and Chern-Simons terms. 
We also outline the procedure for finding the parameters
 of the truncated theory. 
As an example we consider dimensional reduction on $S^2$
 of the 5-dimensional
 minimal supergravity with curvature squared terms
  and obtain the truncated theory
 without any curvature squared terms. This truncated theory 
 reproduces correctly
 the exact central charge of the boundary CFT.

\vfill \eject

\baselineskip=18pt

\tableofcontents

\sectiono{Introduction}

Three dimensional (super-)gravity with negative 
cosmological constant has
played an important role in the study of 
black holes in string theory\cite{9711138,9712251,9901050}.
The theories relevant for string theory however are not theories
of pure
(super-)gravity but (super-)gravity coupled 
to other matter fields containing
higher derivative terms. In the absence of other matter fields the
higher derivative terms in the action can be removed by  field
redefinition and the action may be reduced to the standard
(super-)gravity action whose gravitational part
contains a sum of three terms,
-- the Einstein-Hilbert term, a cosmological constant term and the
Chern-Simons term\cite{wit1,0706.3359}. 
An argument based on AdS/CFT correspondence
suggests that even when matter fields are present one can carry out a
consistent truncation of the theory where only 
(super-)gravity is present, and
action is again that of standard (super-)gravity whose gravitational
sector is
given by the sum of three terms\cite{0705.0735}.
The main ingredient 
of this argument was that in the dual two
dimensional (super-)conformal field theory 
living at the boundary of $AdS_3$
any correlation function
with one
matter field and arbitrary number of (super-)stress tensor 
vanishes, and furthermore
the correlation functions of the (super-)stress 
tensor are determined completely
in terms of the central charge and are independent of the
matter content  of the theory. 
One of the goals of
the present paper is to describe the consistent truncation procedure
directly in the bulk theory without any reference to AdS/CFT
correspondence.  A general
analysis of consistent
truncation to supergravity theory in general dimensions
can be found in \cite{duff,pope,0707.2315}.

Although our analysis is classical, it can in principle 
be applied to
the full quantum effective action.\footnote{If the theory
admits an $AdS_3$ solution we can define the quantum effective
action to be the one whose classical boundary S-matrix reproduces
correctly the full boundary S-matrix of the quantum theory.}
However
in our analysis we shall have to assume
that the initial action is local, \i.e\ is given by an integral of a
local Lagrangian density that admits a derivative expansion. 
Since in general the full quantum effective action can contain
non-local terms, our analysis will not be directly applicable
on these terms.
In
contrast the argument based on AdS/CFT correspondence works
for the full quantum corrected effective action.

After consistent truncation and field redefinition that brings the
action to the standard form, the parameters labelling the action
are the cosmological constant and the coefficient of the Chern-Simons
term. Of them the Chern-Simons term does not change under the
field redefinition required to bring the action to the standard form
but the cosmological constant term is modified. In theories with
extended supersymmetry the cosmological constant can be determined
from the coefficient of a gauge 
Chern-Simons terms\cite{0506176} which also
does not get renormalized 
under the field redefinition; however in
general we need to determine the cosmological constant explicitly.
We describe a simple algebraic procedure 
for determining the cosmological
constant of the final theory in terms of the parameters of the original
action.

Finally we apply our method to the analysis of the three dimensional
gravity that arises from the dimensional reduction on $S^2$
of five
dimensional supergravity with curvature squared 
corrections\cite{0611329}
and calculate the cosmological constant of the final theory after the
field redefinition that brings the action to the standard form.
In this case the theory has a (0,4) supersymmetry and the expected
value of the cosmological constant can be found by relating it to
the coefficient of a gauge Chern-Simons 
term\cite{0506176,0508218}. 
One can also infer it from the results for the black hole entropy
in these theories computed in \cite{0702072,0703087,0703099}.
The result of the explicit calculation
agrees with these predictions.

\sectiono{Field Redefinition of the Bosonic Fields} \label{s2}

In this section we shall describe how the bosonic part of
a (super-)gravity action coupled to matter fields and containing
higher
derivative terms can be brought into the form of a standard
supergravity action via field redefinition and consistent truncation.
We begin with a three dimensional general coordinate invariant
theory of gravity coupled to
an arbitrary set of matter fields. We denote by $g_{\mu\nu}$ the
metric, by $\phi$ the set of all the scalar fields, by $\Sigma$
the set of all other tensor fields, by $R_{\mu\nu}$ the Ricci tensor
associated with the metric $g_{\mu\nu}$ and by $R$ the 
scalar curvature. 
At the level of two derivative terms, the action takes the
form:
\be \label{etwoder}
S_0 + S_{matter}\, ,
\ee
where 
\begin{equation} \label{edefs0}
S_0 = {\int}d^3x \sqrt{-g}\, (R + \Lambda_0(\phi)),
\end{equation}
and $S_{matter}$ denotes the kinetic term for the matter
fields. $-\Lambda_0(\phi)$ represents the scalar field 
potential.
We have
already carried out an appropriate redefinition of the metric to remove
a possible $\phi$ dependent function multiplying $R$ in the 
Einstein-Hilbert term.
If $\Lambda_0(\phi)$ has an extremum at $\phi=\phi_0$
then this theory has a solution where $\phi$ is set equal
to $\phi_0$, all other tensor fields are set to zero,
and the metric is given by that of
an $AdS_3$ space of size $l_0=\sqrt{2/\Lambda_0(\phi_0)}$
for $\Lambda_0(\phi_0)>0$ and a $dS_3$ space of size
$\bar l_0=\sqrt{-2/\Lambda_0(\phi_0)}$ for 
$\Lambda_0(\phi_0)<0$.
In this case $\Lambda_0(\phi_0)$  
corresponds to the {\it negative} of the
cosmological constant.

We shall now consider the effect of adding higher derivative
terms. For this we shall assume that these terms are small
compared to the leading term, in the sense that the length
parameter $l_s$ that controls these terms is small compared
to the length scale $l_0$ over which the leading order solution
varies.\footnote{Often the three dimensional theory is obtained
from dimensional reduction of a higher dimensional theory on a
compact space of size of order $l_0$. In this case if we integrate
out the Kaluza-Klein modes we shall generate higher derivative
terms which are not suppressed by powers of $l_s$. To avoid
this situation we include all the Kaluza-Klein modes in the set
$\Sigma$ without integrating them out.}
We shall also assume that we
can associate with each higher derivative term in the Lagrangian
density
an index $n$ that counts how many powers of $l_s$ accompanies
this term compared to the leading term. For example if the three
dimensional theory is obtained via a dimensional reduction of type
IIB string theory on $K3\times S^1\times S^2\times AdS_3$ with
$K3$ and $S^1$ having size of the order of string scale and $S^2$
and $AdS_3$ having large size, then $\alpha'$ corrections as well
as corrections coming from  integrating out the
heavy modes associated with $K3\times S^1$ compactification
will have index $n>0$, whereas all the terms  associated with
compactification of supergravity on $S^2\times AdS_3$ 
-- including the
ones involving massive Kaluza-Klein modes -- will have
index 0.
An efficient
way to keep track of the derivative expansion is to introduce
a derivative counting parameter $\lambda$ and accompany 
a term of index $n$ by a factor of $\lambda^n$. We shall carry
out our analysis in a power series expansion in $\lambda$ even
though at the end we shall set $\lambda=1$. 

Since in 
three dimension the Riemann tensor $R_{\mu\nu\rho\sigma}$ 
can be expressed
in terms of the Ricci tensor, all the higher derivative terms
can be expressed in terms of the Ricci tensor, its
covariant derivatives and covariant derivatives of the matter fields.
We shall now reorganize these terms as follows.We first note that
under
$g_{\mu\nu}\to g_{\mu\nu} + \delta g_{\mu\nu}$,
\be \label{edeltas}
S_0\to S_0 - \int d^3 x \sqrt{-g}\,  P^{\mu\nu} \delta g_{\mu\nu} 
+O(\delta g^2)\, ,
\ee
where
\be \label{edefp}
P_{\mu\nu}=R_{\mu\nu}-\frac{1}{2}(R 
+\Lambda_0(\phi)) g_{\mu\nu}\, .
\ee
Defining
\be
P\equiv P_\mu^\mu=-\frac{1}{2}R-\frac{3}{2}\Lambda_0(\phi)
\ee
\refb{edefp} can be rewritten as
\be \label{erev}
R_{\mu\nu} = P_{\mu\nu} - 
(P+\Lambda_0(\phi)) g_{\mu\nu}\, .
\ee
We now
eliminate the variables $R_{\mu\nu}, R$ and their covariant 
derivatives in
higher derivative terms by $P_{\mu\nu}, P$ and their 
covariant derivatives.

In this convention the most general action
takes the form:\footnote{During the process of 
replacing $R_{\mu\nu}$ by
the right hand side of \refb{erev}
 we may generate some terms of the form
$\int d^3 x\, 
\sqrt{-g}\, f(\phi)$. Since these cannot be absorbed 
into $\wt S_{matter}$
or $S_n$, we need to absorb them into the scalar field potential
$\Lambda_0(\phi)$ appearing inside $S_0$. Thus $\Lambda_0(\phi)$
needs to be determined in a self-consistent manner. To any order in
power series expansion in $\lambda$ this can be done using an
iterative procedure.}
\begin{equation} \label{egeneral}
S = S_0 +  \lambda\, S_{cs} + \wt S_{matter} +  \lambda^n\, S_n\, .
\end{equation}
$S_0$ is given in \refb{edefs0}. $\lambda\, S_{cs}$
is the gravitational Chern-Simons term
\be \label{edefcs}
S_{cs}=K\, \int d^3 x\, \Omega^{(3)}(\Gamma),
\qquad \Omega^{(3)}(\Gamma)
\equiv \epsilon^{\mu\nu\rho} \left[{1\over 2}
\Gamma^\tau_{\mu\sigma} \p_\nu 
\Gamma^\sigma_{\rho\tau} + {1\over 3}
\Gamma^\tau_{\mu\sigma} \Gamma^\sigma_{\nu\kappa} 
\Gamma^\kappa_{\rho\tau}\right]\, ,
\ee
where $K$ is a constant and $\Gamma^\mu_{\nu\rho}$
denotes the Christoffel symbol. Note that we have included a
factor of $\lambda$ in $S_{cs}$ since in string theory the 
gravitational Chern-Simons term typically arises from
$\alpha'$ corrections.
$\wt S_{matter}$ denotes the matter terms (including the standard
kinetic terms) which  are {\it quadratic and higher order
in $\Sigma$, derivatives of $\Sigma$ and derivatives of $\phi$}.
$\lambda^n\,
S_n$ denotes all other terms, \i.e.\ manifestly general coordinate
invariant terms up to linear order in
$\Sigma$, $\p_\mu\phi$ and their derivatives, but not
terms of the form $\int d^3x\, \sqrt{-g}\, R\, f(\phi)$ since they
can be included in $S_0$.
Most general higher derivative terms in the
action will have the form given in \refb{egeneral}
with $n=1$ 
but for later
use we have allowed
for the fact that the higher derivative terms which cannot be
included in $S_0$, $\wt S_{matter}$ or $\lambda S_{cs}$
may actually begin their expansion at order
$\lambda^n$.
It is easy
to see that $S_n$ must contain 
least one power of $P_{\mu\nu}$, since the $P_{\mu\nu}$
independent terms which do not involve 
$\Sigma$, $\p_\mu\phi$ or their
derivatives can be absorbed into $\Lambda_0(\phi)$ and
$P_{\mu\nu}$
independent terms which 
are linear in $\Sigma$, $\p_\mu\phi$ or their
derivatives either vanish or become 
quadratic in $\Sigma$, $\p_\mu\phi$
or their derivatives after integration 
by parts and hence may be included
in $\wt S_{matter}$. 
 An alert reader may
worry about special cases
where  a symmetric rank 2 tensor $A_{\mu\nu}$
has a coupling proportional to $\sqrt{-g}\, f_1(\phi)\,
g^{\mu\nu} A_{\mu\nu}$ or an
antisymmetric rank three tensor $C_{\mu\nu\rho}$ has a coupling 
proportional to $f_2(\phi)\,
\epsilon^{\mu\nu\rho} C_{\mu\nu\rho}$. We can however avoid
these situations by expressing $A_{\mu\nu}$ as $A g_{\mu\nu}
+ A'_{\mu\nu}$ with $A=g^{\mu\nu}A_{\mu\nu}/3$,
and $A'_{\mu\nu}$ a traceless symmetric matrix,
and $C_{\mu\nu\rho}$ as $C (\sqrt{-g})
\epsilon_{\mu\nu\rho}$ with $C=(\sqrt{-g})^{-1}
\epsilon^{\mu\nu\rho} C_{\mu\nu\rho}/6$, and treating $A$ and $C$
as scalar fields. In this case these terms can be included in the scalar
field potential $\Lambda_0(\phi)$ appearing in $S_0$. Thus
$S_n$ has the form
\begin{equation}
S_n = {\int}d^3x \sqrt{-g} \, P^{\mu\nu}K_{\mu\nu}(\phi,
\Sigma, \nabla_{\rho}, g_{\rho\sigma}, P_{\rho\sigma}, \lambda)\, .
\end{equation}
where $K_{\mu\nu}$ is some combination of 
matter fields, $P_{\mu\nu}$
and their covariant derivatives, and can contain non-negative 
powers
of $\lambda$.

Now consider a redefinition of the metric of the form
\begin{equation}
g_{\mu\nu} \rightarrow g_{\mu\nu} + {\lambda}^nK_{\mu\nu} 
\end{equation}
Under this
\be \label{es0tr}
S_0 \to S_0 -  {\lambda}^n{\int}d^3x\sqrt{-g}\, P^{\mu\nu}K_{\mu\nu} 
+ O(\lambda^{2n}) = S_0 -\lambda^n S_n + O(\lambda^{2n}) \, ,
\ee
\be \label{escstr}
S_{cs}\to S_{cs} + O(\lambda^{n+1})\, ,
\ee
and 
\be \label{esntr}
\lambda^n\, S_n\to \lambda^n\, S_n + O(\lambda^{2n})\, .
\ee
Thus
\be\label{etot}
S_0 +\lambda\, S_{cs} + \lambda^n S_n \to 
S_0 +\lambda\, S_{cs} + O(\lambda^{n+1})\, .
\ee
Furthermore $\wt S_{matter}$   remains
quadratic in $\Sigma$, $\p_\mu\phi$ or their derivatives
under this field redefinition.
The order  $\lambda^{n+1}$ term on the right hand side
of \refb{etot}
can now be regrouped into 
a term of the form $\sqrt{-g}\, f(\phi)$
that can be absorbed into a redefinition of $\Lambda_0(\phi)$,
a term quadratic
in $\Sigma$ and $\p\phi$ that can be absorbed into 
$\wt S_{matter}(\phi)$ and a
term containing at least one power in $P_{\mu\nu}$. Thus the resulting
action may be expressed as:
\begin{equation} \label{enew1}
S = S_0' +  \lambda\, S_{cs} + \wt S_{matter}' +  \lambda^{n+1}\,
S_{n+1},
\end{equation}
where 
\begin{equation}
S_0' = {\int}d^3x \sqrt{-g}\, (R + \Lambda_0'(\phi)),
\end{equation}
$\wt S_{matter}'$ contains terms which  are quadratic and higher order
in $\Sigma$ and derivatives of $\phi$, $\Sigma$ and
\begin{equation}
S_{n+1} = {\int}d^3x \sqrt{-g} P^{\mu\nu}K'_{\mu\nu}(\phi,
\Sigma, \nabla_{\rho}, g_{\rho\sigma}, P_{\rho\sigma}, \lambda)\, 
\end{equation}
for some $K'_{\mu\nu}$.
Thus the new action has the same form as our starting 
action with $n$ replaced by $n+1$.
Repeating this process we can ensure that to any fixed 
order in an expansion in 
$\lambda$, the action can be brought to the form:
\begin{equation} \label{enew2}
S = {\int}d^3x \sqrt{-g}(R + \Lambda(\phi)) +  \lambda
S_{cs} + \wt S_{matter} \, ,
\end{equation}
for some choice of $\Lambda(\phi)$ and $\wt S_{matter}$.

Now suppose $\Lambda(\phi)$ has an extremum at $\phi=\phi_0$. 
Introducing new fields $\xi=\phi-\phi_0$ we may express the
action as
\be \label{efinac}
S = {\int}d^3x \sqrt{-g}(R + \Lambda(\phi_0)) + 
\lambda S_{cs} + \cdots\, ,
\ee
where $\cdots$ contain terms which are at least quadratic in 
$\xi$, $\Sigma$ and their covariant derivatives. We can now carry
out a consistent truncation of the theory by setting $\xi=0$, $\Sigma=0$.
This leaves us with a purely gravitational action with Einstein-Hilbert
term, cosmological constant term and Chern-Simons term.

If the theory contains a 2-form field $B$ with gauge invariance
$B\to B + d\Lambda$ then we can consider a slightly more general
truncation where instead of setting $B$ to zero we set it to have a
constant field strength $C\sqrt{-g} \, \epsilon_{\mu\nu\rho}$ 
for some constant $C$. Let
$\wt B$ denote the fluctuation around this fixed background.
Since $C\sqrt{-g} \, \epsilon_{\mu\nu\rho}$ is a general
coordinate invariant tensor, and since 
the Lagrangian density  depends on $B$ only through the
combination $(dB)_{\mu\nu\rho}=C\sqrt{-g} 
\, \epsilon_{\mu\nu\rho}
+ (d\wt B)_{\mu\nu\rho}$,
it depends on $(d\wt B)_{\mu\nu\rho}$ in
a manifestly general coordinate invariant fashion. We can then proceed
with our analysis as before, including $\wt B$ in the list of tensor
fields $\Sigma$.

If instead of considering a theory of gravity we consider (extended)
supergravity theories, then the theory contains additional
fields. In particular
the additional bosonic fields in the theory are
gauge fields with
Chern-Simons terms\cite{ach1,ach2,9806104,9904010,9904068}. 
Thus in order to show that a general higher
derivative supersymmetric
theory admits a consistent truncation to a supergravity
theory we need to show that higher derivative terms involving
higher powers of gauge fields can be removed by field redefinition.
This follows from the fact that under 
$A_\mu\to A_\mu + \delta A_\mu$ the gauge Chern-Simons term
changes by a term proportional to $\epsilon^{\mu\nu\rho}
Tr\left( F_{\mu\nu} \delta A_\rho\right)$. Thus a term of the form 
$\lambda^n \int \sqrt{-g}\, Tr\left(
F_{\mu\nu} L^{\mu\nu}\right)$ in the
action may be removed (up to order $\lambda^{2n}$ terms) by
a shift of $A_\mu$ proportional to $\sqrt{-g}\,
\epsilon_{\mu\nu\rho}
L^{\nu\rho}$. Following this procedure we can remove
all terms involving the gauge fields other than the Chern-Simons
term to any order in $\lambda$.\footnote{This assumes that all
other terms in the action depend on the gauge field only through
$F_{\mu\nu}$ and not explicitly $A_\mu$, \i.e.\ there are no other
charged fields on the theory. This is not a restriction on the theory
since these charged fields, if present, can be set to zero in a
consistent truncation scheme provided the gauge symmetry is
not spontaneously broken. In the latter case the would be
Goldstone boson associated with the symmetry breaking would mix
with the gauge field via a two point coupling and we cannot have a
consistent truncation to pure supergravity.} 
Once this has been done, one can
then carry out the field redefinition of the metric and the
scalar fields as described earlier, and obtain a consistent
truncation to a theory of metric and gauge fields
with gauge Chern-Simons terms, 
Einstein-Hilbert term, cosmological constant term and
gravitational Chern-Simons term. Supersymmetry then
relates the coefficient of the gauge and gravitational Chern-Simons
terms to the cosmological constant term.

So far our analysis has been restricted to terms in the action
involving bosonic fields only. In a supergravity theory we
must also include the fermionic fields and argue that higher
derivative terms involving the fermions may be removed by
field redefinition. We shall return to this problem in 
\S\ref{sferm}.

\sectiono{Algorithm for Determining $\Lambda(\phi)$} \label{s3}

The analysis of the last section gives an algorithm for
carrying out a field redefinition and consistent truncation
that gives a theory of pure (super-)gravity. However for any
given higher derivative action this is a complicated procedure
and one would like to have a simpler algorithm to 
determine the final truncated theory. Of the various parameters
labelling the final theory the coefficients of the Chern-Simons
terms are easy to determine since they do not get renormalized
from their initial values. On the other hand the cosmological 
constant term does get renormalized during the field redefinition.
In this section we shall
outline a simple procedure for finding the
exact $\Lambda(\phi)$ appearing
in \refb{enew2}
without having to carry out all the steps described
in the last section.
The cosmological constant of the final truncated theory can
then be found by determining the value of $\Lambda(\phi)$
at its extremum.

Suppose our initial action has the form
\begin{equation} \label{e3.1}
S = {\int}d^3x \sqrt{-g}\, \mathscr{L}
+\lambda\, S_{cs}\, .
\end{equation}
In anticipation of the fact that the final truncation involves setting
the scalars $\phi$ to constants and other tensor fields
$\Sigma$ to 0, let us consider a theory of
pure gravity obtained by setting $\Sigma$ to 0 and $\phi$ to some
constant values in \refb{e3.1}. Thus $\phi$ can now be regarded as
a set of external parameters labelling the action. We now 
consider a background
\ben \label{e3.2}
&& ds^2 = -l^2 (1 + r^2) dt^2 + l^2 (1 + r^2)^{-1} dr^2
+ l^2 r^2 d\varphi^2 \, ,\nonumber \\
&& \phi = \hbox{constant}, \qquad \Sigma=0\, ,
\een
representing an $AdS_3$ space of size $l$. If we define
\begin{equation} \label{edeff}
F(l, \phi) = l^3\mathscr{L} 
\end{equation}
evaluated in the background \refb{e3.2},
then the metric satisfies its equation of motion if $l$ is chosen
to be at the extremum $l_{ext}$ of $F$. 
Furthermore $r\, F(l_{ext}, \phi)$ denotes the value
of $\sqrt{-g}\, \LL$ evaluated at the solution. Note that the term
in the equations of motion obtained from the variation of the
Chern-Simons term automatically vanishes for the $AdS_3$ metric
\refb{e3.2} for any constant $l$.

Let us leave this result aside for a while and consider the form of
the action obtained after a field redefinition of the metric
as described in \S\ref{s2}.  After setting $\phi$ to a constant and
$\Sigma$ to 0, the action \refb{enew2} takes the form:
\begin{equation} \label{s3.3}
S = {\int}d^3x\sqrt{-g}(R + \Lambda(\phi)) + \lambda
S_{cs}\, .
\end{equation}
If we evaluate $\sqrt{-g}\, (R + \Lambda(\phi))$  for the 
$AdS_3$ background
\refb{e3.2}, we get a new function $r \, H(l,\phi)$ with
\be \label{e3.4}
H(l,\phi) =  l^3 \left[-\frac{6}{l^2} + \Lambda(\phi)\right]\, .
\ee
Now since we have carried out a field redefinition of the metric but not of
$\Sigma$ or $\phi$, 
we expect $F(l,\phi)$ and $H(l,\phi)$ to be related by a redefinition
of the parameter $l$ for any fixed $\phi$.\footnote{We 
are implicitly
using the result that during the process of redefinition of the metric
the terms arising out of the variation of the Chern-Simons term 
vanishes when the metric has the form 
\refb{e3.2} and $\p_\mu\phi$
and $\Sigma$ are set to zero. This can be seen from the fact that
in this case the field redefinition essentially rescales the metric.
Since $\Gamma^\mu_{\nu\rho}$ remains unchanged under a
rescaling of the metric and since the Chern-Simons term is
constructed entirely in terms of $\Gamma^\mu_{\nu\rho}$, 
it does not change
under such a field redefinition.}
Hence the values of these functions
at the extremum must be the same.
Since the extremum of $H$ 
occurs at,
\be \label{e3.5}
\tilde l_{ext} = \sqrt{\frac{2}{\Lambda(\phi)}}, \qquad 
H(\tilde l_{ext},\phi) = -\sqrt{\frac{32}{\Lambda(\phi)}},
\end{equation}
we get, by setting the right hand side of \refb{e3.5}
to $F(l_{ext},\phi)$,
\begin{equation} \label{e3.6}
\Lambda(\phi) = \frac{32}{F(l_{ext}, \phi)^2}
\end{equation}
provided $F(l_{ext}, \phi)$ is negative. This determines
$\Lambda(\phi)$.

Eq.\refb{e3.6} might give the impression that this procedure
always leads to a theory with positive $\Lambda$, \i.e.\ with
a negative cosmological constant. This is however an artifact
of the fact that we have already assumed that the theory
admits an $AdS_3$ solution. It may so happen that $F(l,\phi)$
defined in \refb{edeff} has an extremum at an imaginary
value of $l$ and hence $F(l,\phi)$ is imaginary at the 
extremum.\footnote{Note that the metric and hence $\LL$ depends
only on $l^2$ and hence is real even when $l$ is imaginary.}
This will give  a negative $\Lambda(\phi)$ and hence  a
positive cosmological constant. A better way to analyze this
case is to consider a de Sitter metric of the form
\be \label{edesi}
ds^2 = -\bar l^2   (1-r^2)dt^2 + \bar l^2 \, (1-r^2)^{-1} \, dr^2
+ \bar l^2 r^2\, d\varphi^2
\ee
instead of the anti-de Sitter metric given in \refb{e3.2}, 
and define
\begin{equation} \label{edefbf}
\bar F(\bar l, \phi) = \bar l^3\mathscr{L}\, ,
\end{equation}
evaluated in this background with $\phi$ set to constants and
$\Sigma$ set to zero.
On the other hand \refb{e3.4} is now replaced by
\be \label{e3.4new}
\bar H(\bar l,\phi) =  \bar l^3 
\left[\frac{6}{\bar l^2} + \Lambda(\phi)\right]\, .
\ee
and the value of $\bar H(\bar l, \phi)$ at the extremum with respect to
$\bar l$ is given by $\sqrt{-32/\Lambda(\phi)}$. 
Equating this to the value of $\bar F$ at its extremum we get:
\be\label{edeflamb}
\Lambda(\phi) = -\frac{32}{\bar F(\bar l_{ext}, \phi)^2}
\end{equation}
provided $\bar F(\bar l_{ext}, \phi)$ is positive.

Finally we note that there is always a possibility that neither
$F(l,\phi)$ nor $\bar F(\bar l,\phi)$ has an extremum for real
values of $l$ or $\bar l$, or even if such extrema exist, the
resulting function $\Lambda(\phi)$ does not have an extremum
as a function of $\phi$. In this case the theory under consideration
does not admit an $AdS_3$ or $dS_3$ solution and we cannot
carry out the consistent truncation following the procedure 
described above.

\sectiono{Higher Derivative Terms Involving 
the Gravitino} \label{sferm}

In the last two sections we have described how via a field
redefinition the bosonic part of the supergravity
action can be brought into the standard form. 
Once the bosonic part of the action has been shown to coincide
with that of the supergravity action one would expect that
supersymmetry will fix the fermionic part of the action
uniquely (up to a possible field redefinition involving the
fermions) to be that of the standard supergravity action. In
this section we shall briefly discuss how such a result
might be proven.

We begin with an action where the purely bosonic part has already
been brought into the standard form using the field redefinition
described in \S\ref{s2}. 
At the onset we shall assume that supersymmetry is unbroken
at the extremum $\phi_0$ of $\Lambda(\phi)$; otherwise
we expect the gravitino to mix with the Goldstino and hence
the matter and the gravity multiplet will 
no longer be decoupled.
This in turn requires $\Lambda(\phi_0)$ to be positive
since we
do not have unbroken supersymmetry in de Sitter space.
If the theory has altogether $\NN$ supersymmetries then there
are $\NN$ gravitino fields $\psi^i_\mu$ with $1\le i\le \NN$.
In the supergravity action of 
\cite{ach1,ach2,9806104,9904010,9904068,0610077} 
the gravitino action
has the form:
\be \label{e7.1}
S^\psi_0 = -\int d^3 x\,\epsilon^{\mu\nu\rho} \bar \psi^i_\mu
\DD_\nu \psi^i_\rho\,,
\ee
where 
\be\label{e7.1a}
\DD_\mu \psi^i_\nu = \p_\mu \psi^i_\nu + {1\over 8} \omega_{ab\mu}
[\gamma^a,\gamma^b] \psi^i_\nu 
\pm \sqrt{\Lambda(\phi_0)\over 32} e_{a\mu}
\gamma^a \psi^i_\nu + A_\mu^a (T^a)_{ij} \psi^j_\nu\, ,
\ee
$\omega_{\mu ab}$ being the spin 
connection, $e_{a\mu}$ the vielbeins,
$A_\mu^a$ the gauge fields and $T^a$ are the generators of the
representation of the gauge group in which the gravitinos transform.
The + ($-$) sign correspond to the gravitinoes associated with
left (right) supersymmetries.
Under a general variation of the gravitino fields
\be \label{e7.2}
\delta S^\psi_0 = -\int d^3 x\,\epsilon^{\mu\nu\rho}\left[
\delta \bar \psi^i_\mu
\DD_\nu \psi^i_\rho + h.c.\right]
\ee
leading to the gravitino equation of motion
\be \label{egr0}
\DD_\nu \psi^i_\rho - \DD_\rho \psi^i_\nu = 0\, .
\ee
The supersymmetry transformation law of the gravitino fields
takes the form
\be \label{e7.3}
\delta_s \psi_\mu^i = \DD_\mu \,\eps^i \,,
\ee
where $\eps^i$ are the supersymmetry transformation parameters.

We shall now examine the possibility of adding 
higher derivative terms in the action and also possibly
in the supersymmetry transformation laws.
Let us denote by $\eta$ the set of all the bosonic and fermionic
fields
coming from the matter sector with the scalars measured
relative to $\phi_0$ (\i.e.\ the set $\eta$ contains the
shifted fields $\xi$
introduced above \refb{efinac}). 
A higher derivative term in the action 
which is quadratic or higher order in
$\eta$ is harmless since we can consistently truncate the
theory by setting $\eta=0$. Thus we need to worry about terms
which are at most linear in $\eta$ or derivatives of $\eta$. 
We shall refer to these
as the dangerous terms since, if present, they will prevent us from
consistently truncating the theory to the one described
by the standard
supergravity action.
As in \S\ref{s2} we shall
organise these  terms 
according to the power of the derivative counting parameter
$\lambda$ that they carry. Let us suppose that
the first dangerous
higher derivative terms in
the Lagrangian density
appear at order $\lambda^k$. Now any term that is proportional
to the equation of motion of the metric, the gauge fields
or the gravitinos derived from the leading supergravity action
can be absorbed into a redefinition of these fields at the
cost of generating higher order terms; thus
we need to look for terms 
which do not vanish identically when leading order supergravity
equations of motion are satisfied. Using this we can remove all the
dangerous terms in the action which contain any power of gauge
field strength, the combination $R_{\mu\nu} + \Lambda(\phi) 
g_{\mu\nu}$, and commutators of covariant derivatives.
Thus the dangerous terms may be expressed as general coordinate 
invariant and local Lorentz invariant combinations of the 
gravitino fields, their symmetrized covariant derivatives and the
metric.
We now consider all 
the order $\lambda^k$ dangerous terms and organise them
by their rank, -- defined as
the total power  of $\psi_\mu$ and $\bar\psi_\mu$
contained in that term.  We begin with the terms of lowest
rank, -- call it $m_0$.
$m_0$ cannot vanish since we have already argued earlier
that all the dangerous terms without the gravitino field can
be removed by field redefinition. (For this we need to include
in the set $\Sigma$ of \S\ref{s2} all the matter fermions as well.)
For non-zero $m_0$ the lowest order
supersymmetry variation of the gravitino described in
\refb{e7.3} has the effect of producing a term of rank 
$(m_0-1)$, constructed out of the gravitino fields, their symmetrized
covariant derivatives, the metric, and covariant derivatives of the
supersymmetry transformation parameter. 
In order for supersymmetry
to be preserved, such terms need to be
cancelled by some other terms. The terms arising from
the supersymmetry variation of the bosons in the original
rank $m_0$ term are of rank $\ge m_0$ and hence cannot cancel
the rank $(m_0-1)$ term.
Thus there are two possibilities: 1) the rank $(m_0-1)$ terms
arising from the variation of the gravitino cancel among themselves
after we integrate by parts and move all the derivatives from
$\epsilon$, $\bar\epsilon$ to the fields, possibly after modifying the
supersymmetry transformation laws of the supergravity
fields,
and 2) we can try to cancel these terms against terms coming
from
supersymmetry variation of the bosons
in a term of rank $(m_0-2)$. Of these the first possibility would
mean that the dangerous terms are invariant under the transformation
\refb{e7.3}
of the gravitino alone up to terms which vanish by lowset order
supergravity equations of motion.\footnote{The terms 
proportional to the lowest order equations of motion of the 
supergravity fields can be
cancelled by modifying the supersymmetry transformation laws
of the supergravity fields, since the additional variation of the
lowest order supergravity action under the modified supersymmetry
transformation laws will be a linear combination of the
lowest order equations of motion of these fields.}
To see if this is possible
we first focus on the terms with maximum number of derivatives
where all the covariant derivatives have been replaced by ordinary
derivatives in the order $\lambda^k$, rank $m_0$
term in the action. The net supersymmetry variation of
these terms under the
supersymmetry transformation law \refb{e7.3} 
must vanish after using the lowest
order gravitino equations of motion \refb{egr0} with $\DD_\mu$
replaced by $\p_\mu$ in \refb{e7.3} and \refb{egr0}, 
since this is the term in $\delta_s S$ with
maximum number of derivatives at this order. In this case the 
gravitino satisfying its lowest order equations of motion has the
form $\psi^i_\mu = \p_\mu \chi^i$, $\bar\psi^i_\mu = 
\p_\mu \bar\chi^i$ for some $\chi^i$, $\bar\chi^i$. 
Let us evaluate the
order $\lambda^k$, rank $m_0$ term in the action in this
background. By assumption the result is not identically zero, -- 
otherwise we could have removed these terms from the action by a
field redefinition of the gravitino field. 
Now for $\psi^i_\mu = \p_\mu \chi^i$, $\bar\psi^i_\mu = 
\p_\mu \bar\chi^i$ the gauge transformation
laws of the gravitino field take the form $\chi^i\to\chi^i+\eps^i$,
$\bar\chi^i\to\bar\chi^i+\bar\eps^i$. Since $\eps^i$ and $\bar\eps^i$ can
be taken to be independent parameters we consider a situation where
only one of the $\eps^i$ is not zero. Invariance under supersymmetry
transformation then tells us that the term under consideration is
invariant under $\chi^i\to \chi^i+\eps^i$ for an arbitrary function
$\eps^i$. In other words the term is independent of $\chi^i$. Repeating
this argument we conclude that the term under consideration must be
independent of all $\chi^i$ and $\bar\chi^i$. Thus it must vanish since
it vanishes when we set all the $\chi^i$ and $\bar\chi^i$ to zero. 
This contradicts our original assertion that the
term does not vanish identically.  This leads us to the
conclusion that the original order $\lambda^k$, rank $m_0$
term in the action, with covariant derivatives
replaced by ordinary derivatives, must have been such that after
suitable integration by parts and commutation of the derivative
operators it
vanishes when the gravitino satisfies its lowest order equation of
motion.

How does the conclusion change when the ordinary
derivatives are replaced by covariant derivatives? 
Since we know that the term can be 
manipulated and shown to vanish
when covariant derivatives are replaced 
by ordinary derivatives, we
can carry out the same manipulation.
The only possible
extra terms which could arise must be proportional to 
the commutators $[D_\mu,D_\nu]$ since the covariant derivatives
can be manipulated in the same manner as the ordinary derivatives
except for their commutators. However these commutators can
be reduced to terms with lower number of derivatives using
the lowest order metric and gauge field equations
of motion. We can now repeat our analysis on these left-over
terms with lower number of derivatives and show that they must
be further reducible to terms with lower number of derivatives.
Repeating this procedure we can show 
that a term that is invariant under the lowest
order supersymmetry
transformation of the gravitino alone, must vanish as a 
consequence of lowest order supergravity field equations, and hence
can be removed by a field redefinition.

We now turn to the second possibility. This
requires the action to
contain higher derivative terms of order $\lambda^k$ and rank
$(m_0-2)$. Since 
by assumption the action does not contain
any dangerous term of rank $(m_0-2)$ to order $\lambda^k$,
the only possibility is to try to generate these terms from the
supersymmetry variation of a non-dangerous term of rank $(m_0-2)$.
In order to rule out this possibility we need to make one assumption:
{\it as
a consequence of unbroken supersymmetry 
the matter sector fields transform to terms which contain
at least a single power of the matter sector field,} \i.e.\ we  have
$\delta_s \eta \sim  \OO(\eta)$.\footnote{This is of course
true at the lowest order in $\lambda$ but we shall assume that
this property continues to hold even after including possible
higher derivative corrections to the supersymmetry transformation
laws.}
In this case terms quadratic and higher order in $\eta$ transform
to terms quadratic and higher order in $\eta$ and cannot cancel
terms which are at most linear in $\eta$. This rules out the last
possibility. Thus we see that it is not possible to add higher derivative
dangerous terms in the action in a manner consistent with
supersymmetry.

\sectiono{Dimensional Reduction of
Five Dimensional Supergravity}

In this section we shall consider five dimensional 
supergravity with curvature squared term coupled to a set of
vector multiplets\cite{0611329} and dimensionally 
reduce this theory on $S^2$ in the presence of background
magnetic flux through $S^2$ to get a three 
dimensional (0,4) supergravity with curvature squared term,
coupled to a set of matter fields.
We then apply the procedure of \S\ref{s2} and \S\ref{s3} to
truncate this to a pure supergravity theory with 
gravitational
Chern-Simons term, but no other higher derivative terms.

We shall concentrate our attention on the part of the action
involving the bosonic fields only. In the three dimensional
theory this involves the metric and an SU(2) gauge field that
arises during the dimensional reduction of the five dimensional
theory on $S^2$.
As we have seen at the end of \S\ref{s2}, reducing the gauge field
action to pure Chern-Simons term is relatively simple; hence we
shall focus on the part of the action involving the 
metric.
For
this we can restrict the fields to the SU(2) invariant sector
from the beginning.
Since the SU(2) R-symmetry of the three dimensional 
supergravity can be identified with the rotational symmetry of
the compact $S^2$, this allows us to carry out the dimensional
reduction by restricting the field configurations to
rotationally invariant form.\footnote{One might worry 
about the extra terms which
may be generated during the redefinition of the gauge field
that brings the gauge field action into the standard form; however
one can easily argue that these terms cannot affect the final form
of the action involving the metric since setting all $SU(2)$
non-invariant fields, including the gauge fields, to zero 
provides a
consistent truncation of the theory.}

The five dimensional $\NN=2$ supergravity
has a Weyl multiplet, a set of vector multiplets and
 a compensator hypermultiplet. After gauge fixing to
 Poincare supergravity,  
 the bosonic fields of the theory include
the metric 
$g_{ab}$, the two-form auxiliary field
$v_{ab}$, a scalar auxiliary field $D$, 
a certain number ($n_V$) of one-form gauge fields $A^I_a$
with $1\le I\le n_V$, and an equal number of scalars
$M^I$ \cite{0611329}.
Here  $a,b,..$ are five dimensional coordinate labels and run from
0 to 4.
We shall denote by $F^I =
dA^I$ the field strength associated with the gauge field $A^I$.
The
action for bosonic fields including curvature squared terms can be
written as
\begin{equation} \label{es1}
 S = \frac{1}{4{\pi}^2}\int d^5x\sqrt{-g^{(5)}}[\mathscr{L}_0 + \mathscr{L}_1]
\end{equation}
where  $\mathscr{L}_0$ is the lagrangian at two derivative order and
$\mathscr{L}_1$ denotes the supersymmetric completion of the
curvature squared terms. 
The explicit forms of $\LL_0$ and $\LL_1$  are\cite{0611329,0703087}
\begin{equation}
\begin{split}
\mathscr{L}_0
 &=-2 \left(\frac{1}{4}D - \frac{3}{8}R 
 - \frac{1}{2}v^2\right) 
 + N\left(\frac{1}{2}D + \frac{1}{4}R +3v^2\right) 
+ 2N_Iv^{ab}F^I_{ab}\\
  &+N_{IJ}\left(\frac{1}{4}F^I_{ab}F^{Jab} 
  + \frac{1}{2}
  \partial_aM^I\partial^aM^J\right) + \frac{1}{24}
  e^{-1}c_{IJK}A^I_aF^J_{bc}F^K_{de}\epsilon^{abcde}
\end{split}
\end{equation}
\begin{equation}
\begin{split}
\mathscr{L}_1= \frac{c_{2I}}{24} &\biggl[\frac{1}{16}e^{-1}
\epsilon_{abcde}A^{Ia}C^{bcfg}C^{de}_{fg}
+ \frac{1}{8}M^IC^{abcd}C_{abcd} +\frac{1}{12}M^ID^2 +
\frac{1}{6}F^{Iab}v_{ab}D \\
&-\frac{1}{3}M^IC_{abcd}v^{ab}v^{cd} 
- \frac{1}{2}F^{Iab}C_{abcd}v^{cd}
+\frac{4}{3}M^I\nabla^av^{bc}\nabla_av_{bc}
+ \frac{4}{3}M^I{\nabla}^av^{bc}{\nabla}_bv_{ca} \\ 
&+ \frac{8}{3}M^I
\left(v_{ab}\nabla^b\nabla_cv^{ac} +
\frac{2}{3}v^{ac}v_{cb}R_a^b + \frac{1}{12}
v^{ab}v_{ab}R\right)
-\frac{2}{3}e^{-1}M^I
\epsilon_{abcde}v^{ab}v^{cd}\nabla_fv^{ef}\\ 
&+\frac{2}{3}e^{-1}F^{Iab}\epsilon_{abcde}
v^{cf}\nabla_fv^{de}+e^{-1}F^{Iab}
\epsilon_{abcde}v^c_f\nabla^dv^{ef}-
\frac{4}{3}F^{Iab}v_{ac}v^{cd}v_{db} \\
&- \frac{1}{3}F^{Iab}v_{ab}v^2 +4M^Iv_{ab}v^{bc}v_{cd}v^{da} -
M^I(v_{ab}v^{ab})^2 \biggr]
\end{split}
\end{equation}
where $c_{IJK}$ and $c_{2I}$ are parameters of the
theory,  $e\equiv\sqrt{-g}$, and
\begin{eqnarray}
N = \frac{1}{6}c_{IJK}M^IM^JM^K\\
N_I = \frac{1}{2}c_{IJK}M^JM^K\\
N_{IJ} = c_{IJK}M^K\, ,
\end{eqnarray}
and $C_{abcd}$ is the Weyl tensor defined as
\begin{equation}
C^{ab}_{cd}= R^{ab}_{cd} + \frac{1}{6}
R\delta^{[a}_{[c}\delta^{b]}_{d]} 
- \frac{4}{3}\delta^{[a}_{[c}R^{b]}_{d]}\, .
\end{equation}
The parameters $c_{2I}$ appear in the coefficients of the
higher derivative terms; thus we can keep track of the derivative
expansion by simply counting the power of $c_{2I}$ appearing in
the various terms.

We now carry out the
dimensional reduction on $S^2$ and focus on the sector
invariant under the $SO(3)$ isometry group of $S^2$.
This can be done using the following ansatz for the five
dimensional fields
\begin{equation} \label{edim}
\begin{split}
&ds^2 = g^{(3)}_{\mu\nu}(x) 
dx^{\mu}dx^{\nu} + {\chi}^2(x)d{\Omega}^2, \qquad
0\le\mu,\nu\le 2\\
&v_{\theta\phi} = V(x)\sin\theta\\
&F^I_{\theta\phi} = \frac{p^I}{2}\sin\theta, \qquad
F^I_{\mu\nu} = \p_\mu A^I_\nu - \p_\nu A^I_\mu\, ,
\end{split}
\end{equation}
with the mixed components of $F^I_{ab}$ and $v_{ab}$ set
to zero. 
Here $x^\mu$  denote
the three dimensional coordinates. All the scalar fields can
be arbitrary functions of $x$ but are independent of the
coordinates $(\theta,\phi)$ of $S^2$. For the metric given in
\refb{edim} the non-vanishing components of the Riemann
tensor are
\ben \label{enonvan}
&& R_{\mu\nu\sigma\rho}=R^{(3)}_{\mu\nu\sigma\rho},\quad
R_{i\mu j\nu} = -\chi^{-1}\, g_{ij} \, \nabla_\mu\nabla_\nu \chi,
\quad R_{ijkl} = \chi^{-2} \left(g_{ik} g_{jl} - g_{il} g_{jk}
\right)
\left(1 - g^{(3)\mu\nu}\p_\mu\chi\p_\nu\chi\right)\, ,\nonumber \\
&& \qquad \qquad 0\le\mu,\nu \le 2, \qquad i,j=\theta,\phi\, .
\een
Here $R^{(3)}_{\mu\nu\rho\sigma}$ is the Riemann tensor
and $\nabla_\mu$ is the covariant derivative
computed using the three dimensional metric $g^{(3)}_{\mu\nu}$.
Using these relations we get the dimensionally
reduced action to be
\begin{equation} \label{es2}
\begin{split}
S = &-\frac{c_2\cdot p}{96\pi}\int d^3x 
\Omega^{(3)}({\Gamma}) \\&+ \int
d^3x\sqrt{-g^{(3)}}\frac{\chi^2}{\pi}\biggl(\frac{3}{4}
+
\frac{1}{4}N + \frac{c_2\cdot M}{288}\frac{1}{\chi^2} 
+ \frac{c_2\cdot
  M}{72}\frac{V^2}{\chi^4} - \frac{c_2\cdot
  p}{288}\frac{V}{\chi^4}\biggr)R^{(3)}\\
&+\int d^3x \sqrt{-g^{(3)}}\frac{\chi^2}{\pi}U(\chi, M^I, V, p^I, D) \\
&+\int d^3x \sqrt{-g^{(3)}} \frac{\chi^2}{\pi}\frac{c_2\cdot
  M}{192}\biggl(\frac{8}{3}R^{(3)}_{\mu\nu}R^{(3)\mu\nu} -
\frac{5}{6}R^{(3)2} +
\frac{16}{3\chi}R^{(3)}_{\mu\nu}\nabla^{\mu}\nabla^{\nu}\chi -
\frac{4}{3\chi}R^{(3)}\nabla^2\chi\biggr) \\
&+ \int d^3x \sqrt{-g^{(3)}}\wh{\mathscr{L}}(\chi, v_{\mu\nu}, M^I, 
F^I_{\mu\nu}, R^{(3)}_{\mu\nu})
\end{split}
\end{equation}
where
\begin{equation}
\begin{split}
U(\chi, M^I, V, p^I, D) = &\frac{2}{\chi^2}
\biggl(\frac{3}{4}+
\frac{1}{4}N\biggr) -2\biggl(\frac{1}{4}D
-\frac{V^2}{\chi^4}\biggr) 
+ N\biggl(\frac{1}{2}D + \frac{6V^2}{\chi^4}\biggr)\\
& + \frac{2(N\cdot p)V}{\chi^4} +
\frac{N_{IJ}p^Ip^J}{8\chi^4} 
+ \frac{c_2\cdot M}{96\chi^4} +\frac{c_2\cdot
  M}{288}D^2 + \frac{c_2\cdot p}{144}\frac{VD}{\chi^4} \\
&- \frac{5}{36}(c_2\cdot M)\frac{V^2}{\chi^6} - \frac{c_2\cdot
    p}{48}\frac{V}{\chi^6} 
    + \frac{c_2\cdot p}{36}\frac{V^3}{\chi^8} 
    + \frac{c_2\cdot M}{6}
    \frac{V^4}{\chi^8}\\
\end{split}
\end{equation}
and $\wh{\mathscr{L}}(\chi, v_{\mu\nu}, M^I, F^I_{\mu\nu},
R^{(3)}_{\mu\nu})$ denotes terms which are at least
quadratic in $\nabla_\mu\chi, v_{\mu\nu}, \nabla_\mu 
M^I$ and $F^I_{\mu\nu}$. In eq.\refb{es2} all covariant derivatives
are computed using the three dimensional metric $g^{(3)}_{\mu\nu}$.

We first need to redefine our metric in 
such a manner that the coefficient of
$R^{(3)}$ in the second line of the action 
\refb{es2} can be absorbed into the metric.
We define 
\begin{equation} \label{emetricdef}
\wt g_{\mu\nu} = \psi^{-2} \, {g}^{(3)}_{\mu\nu}
\end{equation}
where
\begin{equation}
\psi^{-1} = \frac{\chi^2}{\pi}\biggl(\frac{3}{4} +
\frac{1}{4}N + \frac{c_2\cdot M}
{288}\frac{1}{\chi^2} + \frac{c_2\cdot
  M}{72}\frac{V^2}{\chi^4} - \frac{c_2\cdot
  p}{288}\frac{V}{\chi^4}\biggr)
\end{equation}
After substituting \refb{emetricdef}
into the action \refb{es2},
we get 
\begin{equation} \label{eaction3}
\begin{split}
S =  &-\frac{c_2\cdot p}{96\pi}\int d^3x 
\Omega^{(3)}(\wt{\Gamma}) \\
& + \int d^3x \sqrt{-\wt g}\biggl[\wt R 
+ Z(\chi, M^I, V, p^I, D)\biggr]\\
&+ \int d^3x\sqrt{-\wt g}\frac{\chi^2}{\pi\psi}\frac{c_2\cdot
  M}{192}\biggl(\frac{8}{3}\wt R^{\mu\nu}\wt R_{\mu\nu} -
\frac{16}{3\psi} \wt R^{\mu\nu}\wt \nabla_\mu\wt \nabla_\nu\psi +
\frac{4}{3\psi}\wt R\wt \nabla^2\psi - \frac{5}{6}\wt R^2 +
\frac{16}{3\chi}\wt R^{\mu\nu}\wt \nabla_\mu\wt \nabla_\nu\chi -
\frac{4}{3\chi}\wt R\wt \nabla^2\chi\biggr)\\ 
 &+ \int d^3x \sqrt{-\wt g}\mathscr{\wh L'}(\chi, v_{\mu\nu}, M^I,
F^I_{\mu\nu},\wt {R}_{\mu\nu})
\end{split}
\end{equation}
where
\begin{equation}
 Z(\chi, M^I, V, p^I, D) = \psi^3 \frac{\chi^2}{\pi}U(\chi, M^I, V, p^I, D)
\end{equation}
and $\wh \LL'$ denotes terms quadratic and higher order in the derivatives
of scalar fields and other tensor fields.
For shorthand notation we denote all scalar fields by 
$\phi$ i.e.$(\chi, M^I,
V, p^I, D)\equiv \phi$.

Following the general procedure given in \S\ref{s2} we now
define
\begin{equation}
\begin{split}
&P_{\mu\nu} = \wt R_{\mu\nu} - \frac{1}{2}\wt {g}_{\mu\nu}[\wt R +
 \Lambda_0(\phi)]\\
&P = -\frac{1}{2}\wt R - \frac{3}{2}\Lambda_0(\phi)\, ,
\end{split}
\end{equation}
where $\Lambda_0(\phi)$ is a function to be determined later, 
and rewrite the action as
\begin{equation} \label{es3pre}
\begin{split}
S =  &-\frac{c_2\cdot p}{96\pi}\int d^3x 
\Omega^{(3)}(\wt{\Gamma}) + \int d^3x \sqrt{-\wt g}\biggl[\wt R 
+ Z(\phi)\biggr] + \int d^3x
\sqrt{-\wt g}
P_{\mu\nu}K^{\mu\nu}\\
& + \int d^3x \sqrt{-\wt g}\frac{\chi^2}
{\psi\pi}\frac{c_2\cdot M}{384} \Lambda_0^2(\phi)\\
& + \int d^3x \sqrt{-\wt g}\, \mathscr{\wt L}\end{split}
\end{equation}
where
\begin{equation}
\begin{split}
K_{\mu\nu} =& \frac{\chi^2}{\psi\pi}\frac{c_2\cdot M}{192}\bigg[
\frac{8}{3}P_{\mu\nu} - \frac{2}{3}\wt g_{\mu\nu}P
+\frac{2}{3}\wt g_{\mu\nu}\Lambda_0(\phi) 
- \frac{16}{3\psi}\wt \nabla_\mu \wt
\nabla_\nu\psi \\&+ \frac{8}{3\psi}\wt g_{\mu\nu}\wt \nabla^2\psi +
\frac{16}{3\chi}\wt \nabla_\mu\wt \nabla_\nu \chi -
\frac{8}{3\chi}\wt g_{\mu\nu}\wt \nabla^2\chi\bigg]\, ,
\end{split}
\end{equation}
and $\wt \LL$ denotes terms quadratic and higher order in the derivatives
of the scalar fields and other tensor fields.
We now choose $\Lambda_0(\phi)$ to be the solution to the equation
\be\label{elaeq}
\Lambda_0(\phi) = Z(\phi) + \frac{\chi^2}
{\psi\pi}\frac{c_2\cdot M}{384} \Lambda_0(\phi)^2\, ,
\ee
so that the action \refb{es3pre} may be expressed as
\begin{equation} \label{es3}
\begin{split}
S =  &-\frac{c_2\cdot p}{96\pi}\int d^3x 
\Omega^{(3)}(\wt{\Gamma}) + \int d^3x \sqrt{-\wt g}\biggl[\wt R 
+ \Lambda_0(\phi)\biggr] + \int d^3x
\sqrt{-\wt g}
P_{\mu\nu}K^{\mu\nu}\\
& + \int d^3x \sqrt{-\wt g}\, \mathscr{\wt L}\, .
\end{split}
\end{equation}
In this case, as we mentioned earlier, the required field
redefinition which will remove the 
four derivative terms from the action
\refb{es3} is 
\begin{equation}
\wt g_{\mu\nu} \rightarrow \wt g_{\mu\nu} +
K_{\mu\nu}\,.
\end{equation}
To this order the scalar field potential $-\Lambda(\phi)$ is
given by
\be \label{escalar}
\Lambda(\phi) = \Lambda_0(\phi)=
Z(\phi) + \frac{\chi^2}
{\psi\pi}\frac{c_2\cdot M}{384} Z^2(\phi) + \OO(c_2^2)\, .
\ee

This process can now be repeated to remove the six 
and higher derivative
terms from the action, but we shall not go through the details
of the analysis. Our interest is
in finding the exact expression 
for $\Lambda(\phi)$ since this is what controls the final
truncated action.
We have
already described the algotithm for finding 
$\Lambda(\phi)$ 
in \S\ref{s3}. The first step is to compute $F(l,\phi)$ for the action
\refb{eaction3} by evaluating the Lagrangian density (without the
Chern-Simons term) in the $AdS_3$ background \refb{e3.2}
with constant scalar fields and vanishing tensor fields.
We get
\begin{equation}
F(l, \phi) = -6l+ l^3Z(\phi) 
+ 2a\frac{1}{l}
\end{equation}
where 
\begin{equation}
a = \frac{\chi^2}{\psi\pi}\frac{c_2\cdot M}{192}\, .
\end{equation}
The extremum of $F(l,\phi)$ with respect to $l$ 
occurs at\footnote{There is, in principle, another extremum at
$l_{ext}^2=(Z(\phi))^{-1} \, \left(1 - \sqrt{1 + 2 a Z(\phi)/3}
\right)$. This could
in principle describe a de Sitter solution. However since for this
solution $|l_{ext}|\sim a$, the radius is small and there is no
systematic derivative expansion.}
\begin{equation}
l_{ext}^2 = \frac{1}{Z(\phi)} + \frac{1}{Z(\phi)}
\sqrt{1 + \frac{2a}{3}Z(\phi)}\, .
\end{equation}
Hence $\Lambda(\phi)$ is given by
\begin{equation} \label{efull}
\Lambda(\phi) ={32\over F(l_{ext},\phi)^2}= 
{32 Z(\phi)\over W(\phi)} \left( 2a {Z(\phi)\over W(\phi)}
+ W(\phi) - 6\right)^{-2}\, , \qquad
W(\phi) \equiv 1 + \sqrt{1 + \frac{2a}{3}Z(\phi)}\, .
\end{equation}
Before we proceed we note that to order $c_{2I}$ terms, \i.e.\
order $a$ term, eq.\refb{efull} reduces to
\be \label{ecomp}
\Lambda(\phi) = Z(\phi) + {1\over 2} \, a \, Z(\phi)^2 + \OO(a^2)\, .
\ee
This agrees with the result \refb{escalar}
of the explicit calculation to this order.

We now return to the full expression \refb{efull} for
$\Lambda(\phi)$.
$\Lambda(\phi)$ has an extremum at the
supersymmetric attractor point\cite{0702072,0703087}
\begin{equation}
\begin{split}
&\chi= \frac{pb}{2}\\
&M^I = \frac{p^I}{pb}\\
&V = -\frac{3pb}{8}\\
&D = \frac{12}{p^2b^2}\\
\end{split}
\end{equation}
where
\begin{equation}
p^3\equiv {1\over 6} c_{IJK} p^I p^J p^K,
\qquad b^3 = 1 + \frac{c_2\cdot p}{12p^3} 
\end{equation}
The value of $\Lambda(\phi)$ at it's extremum is given by
\begin{equation}
\Lambda(\phi_0) 
= \frac{32\pi^2}{p^6}\biggl[1 + \frac{c_2\cdot p}{8p^3}\biggr]^{-2} 
\end{equation}
Thus the final truncated theory, obtained by setting 
$\phi$ to its value at the
extremum and other matter fields to zero, is given by
\be \label{efin}
S = \int d^3 x \, \sqrt{-\wt g} \, (\wt R + \Lambda(\phi_0))
-\frac{c_2\cdot p}{96\pi}\int d^3x 
\Omega^{(3)}(\wt{\Gamma})\, .
\ee
{}From this one can compute 
the central charges of the conformal field theory  
living on boundary of AdS using standard formul\ae\ (see
{\it e.g.} \cite{0705.0735}). The result
is 
\begin{equation}
\begin{split}
&c_L = 24\pi\biggl(\sqrt{\frac{2}{\Lambda(\phi_0)}} - \frac{c_2\cdot
  p}{96\pi}\biggr) = 6p^3 + \frac{1}{2}c_2\cdot p\\
&c_R =  24\pi\biggl(\sqrt{\frac{2}{\Lambda(\phi_0)}} + \frac{c_2\cdot
  p}{96\pi}\biggr) = 6p^3 + c_2\cdot p
\end{split}
\end{equation}
These results agree with the predictions of \cite{0506176,0508218}
from the requirement of (0,4) supersymmetry, as well as the 
explicit calculations of \cite{0702072,0703087,0703099} 
from the computation of the black
hole entropy.

\end{document}